\documentclass[aps,pra,reprint,superscriptaddress,floatfix,showpacs]{revtex4-1}
\usepackage{blindtext}
\usepackage[utf8]{inputenc}
\usepackage[T1]{fontenc}
\usepackage{bm}
\usepackage{dcolumn}
\usepackage{graphicx}
\usepackage{subfigure}
\begin{document}
\title{Electroplating based engineering of plasmonic nanorod metamaterials for biosensing applications}

\author{Mihir Kumar Sahoo}
\thanks{These two authors contributed equally}
\author{Abhay Anand VS}
\thanks{These two authors contributed equally}
\author{Anshuman Kumar}
\email{anshuman.kumar@iitb.ac.in}
\affiliation{Laboratory of Optics of Quantum Materials (LOQM), Department of Physics, IIT Bombay, Mumbai, 400076, Maharashtra, India}

%\keywords{Electroplating; Gold nanorod; Hyperbolic metamaterial; Plasmonics; Optical biosensor}

\begin{abstract}
Sensing lower molecular weight in a diluted solution using a label-free biosensor is challenging and requires a miniaturized plasmonic structure, e.g., a vertical Au nanorod (AuNR) array based metamaterials. The sensitivity of a sensor mainly depends on transducer properties and hence for instance, the AuNR array geometry requires optimization. Physical vapour deposition methods (e.g., sputtering and e-beam evaporation) require a vacuum environment to deposit Au, which is costly, time-consuming, and thickness-limited. On the other hand, chemical deposition, i.e., electroplating deposit higher thickness in less time and at lower cost, becomes an alternative method for Au deposition. In this work, we present a detailed optimization for electroplating based fabrication of these metamaterials. We find that slightly acidic (6.0 < pH < 7.0) gold sulfite solution supports immersion deposition, which should be minimized to avoid uncontrolled Au deposition. Immersion deposition leads to plate-like (for smaller radius AuNR) or capped-like, i.e., mushroom (for higher radius AuNR) structure formation. The electroplating time and DC supply are the tuning parameters that decide the geometry of the vertically aligned AuNR array in area-dependent electroplating deposition. This work will have implications for developing plasmonic metamaterial based sensors.

\end{abstract}
%\begin{document}

%\flushbottom
%\maketitle
%\thispagestyle{empty}

\maketitle
\section{Introduction}

Biological-sensor or biosensor senses the properties of biomolecule present in an analyte, e.g., concentration of the substance, through a transducer that converts one form of energy to another\cite{Bhalla2016}. Biosensors may be classified as first-generation, second-generation, and third-generation biosensors\cite{2005}. The first-generation biosensor is based on electrical response, i.e., current or voltage response of the biomolecule. In electrical biosensors, the transducer produces a low and superimposed electrical signal in an ambient atmosphere, which requires signal processing, i.e., subtracting reference signal followed by amplifying the resultant signal. The analog signal is usually converted to a digital signal and passed through a microprocessor stage to process the data and convert it to concentration units, which may not sense for lower molecular weight. Second-generation biosensors use a foreign molecule as a mediator, chemically or temporarily attached, to detect molecular presence or activity through a labeling process. The mediator's response to the applied signal is analyzed to detect the properties of the biomolecule, also known as a label-based biosensor. However, in label-based biosensors, the mediator (foreign molecule) can alter the binding properties of the biomolecule, e.g., antibody-antigen interaction, which introduces systematic error to the biosensor analysis. Therefore, scientists are more interested in the third-generation biosensor, where reaction of the biomolecule to the input signal produces a response without the mediator, i.e., label-free biosensors, using metamaterial.

Depending on the signal transduction, biosensors can be classified broadly into five types\cite{Damborsk2016}, such as (i) optical, (ii) electrochemical, (iii) thermometric, (iv) piezoelectric, and (v) magnetic. Optical label-free biosensor, which produces a signal proportional to the concentration of measured analyte, exhibits various advantages, such as high specificity, high sensitivity, small size, and cost-effectiveness. The optical biosensor measures change in absorption/reflection/transmission of light when a chemical reaction occurs or the quantity of light emitted by some luminescent process.

Shining light on the analyte contained in any substrate may not detect the biomolecule due to lower sensitivity. In particular, metal substrates (e.g., Na, In, Al, Cu, Ag, and Au) are preferred over others due to their surface plasmon resonance (SPR) property that allows free surface electrons to oscillate collectively while shined by a specific wavelength of light. The Na metal is highly reactive, whereas the In is expensive to use in the biosensor. The Al, Cu, and Ag are susceptible to oxidation; therefore, as a traditional SPR precious metal, Au exhibits better oxidation and corrosion resistance, improving the electron transfer efficiency\cite{Si2021}. However, the bare Au substrates are unsuitable for the biosensor because of the poor absorbance properties of biomolecules\cite{Wu2010}, i.e., lower polarizability issue while detecting lower molecular weight (< 500 g/mol) biomolecules in a highly diluted solution. 

Miniaturization of Au substrate opens up two new horizons, i.e., (i) localized surface plasmon resonance (LSPR)\cite{Baqir2020} and (ii) hyperbolic metamaterial (HMM) property\cite{Chen2012,Sreekanth2016}, over a broad wavelength, ranging from visible to near-infrared (NIR) region. At resonance, i.e., matching of LSPR wavelength and incoming electromagnetic wavelength, the light is absorbed in the nanostructured Au substrate depending on its polarizability. The basic principle of the plasmonic sensing mechanism is the excitation of charge density oscillations (surface plasmons) propagating along the metal-dielectric interface when the wavevector of incident light satisfies the resonant condition\cite{Kim2006}. At visible and NIR wavelengths, the electric field associated (decays exponentially) with these oscillations is highly sensitive to the change in the refractive index of its surrounding medium. The plasmonic sensor facilitates real-time monitoring of these biomolecular binding events\cite{Smith2015,FelixRendon2021}. The HMM property amplifies the evanescent waves and characteristic sub-wavelength resolution that enhances sensitivity.

The polarizability depends on various factors, such as shape, size, material composition, and surrounding medium of the Au nanostructure. Chen et al.\cite{Chen2008} reported sensitivity of various sizes and shapes of Au nanostructures (nanospheres, nano-cubes, nanorods, nano-bipyramids, and nano-branches) by dispersing them in a water-glycerol solution. The Au nanosphere gives the lowest sensitivity, while the Au nano-branches show high sensitivity toward refractive index change. However, reproducibility is a vital issue while fabricating the Au nano-branches. Definite orientation and reproducible Au nanostructures is still an unexplored research area to investigate refractive index sensitivity. Among others, the geometry of Au nanorod (AuNR) can be tuned (by length and diameter) minutely compared to nanosphere structure, where a radius is the only tuning parameter. Besides, the AuNR exhibits a higher bulk refractive index sensitivity and a narrower line width than other geometries that determine accurate peak shifts\cite{Li2015,Liu2015}. Tuersun et al.\cite{Tuersun2017} reported that Au nanocylinder exhibits a higher refractive index sensitivity than Au nanoellipsoid, Au nanobar, and Au nanocone\cite{Liu2015}.

The AuNR can be horizontal\cite{Hwang2022} or vertical\cite{Yanagawa2019}, producing transverse SPR and longitudinal SPR wavelength, respectively\cite{Lohse2013,Nguyen2016}. The longitudinal SPR (vertical AuNR) exhibits higher sensitivity towards aspect ratio (length/diameter) change than the transverse SPR. However, the SPR occurs at lower energy or higher wavelength for easily polarized longitudinal AuNR. The definite order of the vertical AuNR array can bring down the SPR to a lower wavelength\cite{Kabashin2009}, i.e., in visible and NIR regions. In the vertical AuNR array, the tuning parameters for SPR are (i) filling ratio (depends on radius and periodicity), (ii) length, and (iii) surrounding medium. By controlling the geometry of vertical AuNR, i.e., radius, periodicity, and length, the SPR can be tuned. To operate the optical biosensor in the visible region (i.e., LSPR wavelength), the feature size of the AuNR array must be less than the visible wavelength\cite{Petryayeva2011}. Palermo et al.\cite{Palermo2020} reported that when the distance between the nanorods is smaller than the wavelength, the HMM supports a guided mode with the field distribution inside the layer determined by plasmonic-mediated interaction between nanorods. This anisotropic guided mode has resonant excitation conditions similar to the surface plasmon polariton mode of a smooth metal film with a considerable probe depth (500 nm)\cite{Palermo2020}. The AuNR array is fascinating as they exhibit anisotropic optical properties throughout the visible and NIR spectral range, a property related to LSPR across their long and short axes. Vasilantonakis et al.\cite{Vasilantonakis2015} reported that the diameter, periodicity (filling ratio < 0.6), and length must be less than 400 nm each to get an optimized optical sensitivity in the visible region. Researchers reported that the strongest field enhancement occurs close to the edge or in the gaps between the NRs. 

Electron beam lithography (EBL) is an effective technique for fabricating low-dimensional, reproducible nanostructures. Therefore, vertical cylindrical AuNR array fabrication is carried out via EBL, followed by metallization. Among various physical vapour deposition (PVD) techniques sputtering and e-beam evaporation technique are popular due to smooth and uniform thin film deposition of Au. However, the lower deposition rate and heating up of the instrument limit the PVD to deposit higher thickness. On the contrary, chemical deposition methods, e.g., electroplating, surpass PVD's limitations to depositing the AuNR array in the ambient atmosphere at a higher deposition rate. An Au electroplating method uses a chemical solution containing Au, such as potassium gold cyanide, gold chloride, and gold sulfite. Cyanide is a highly toxic and surface-active chemical that is difficult to rinse, giving rise to staining problems. The gold chloride solution is acidic, with a pH value between 2.0 and 6.0, allowing immersion and electrochemical deposition. Controlling the thickness in immersion deposition is a tiresome and challenging task. A slightly acidic or neutral gold sulfite (pH: 6.0-7.0) solution deposits a uniform Au layer on the seed metal layer, minimizing the immersion deposition. This work considers sputtering, e-beam evaporation, and electroplating techniques to optimize the vertical AuNR array—deposition of smooth and uniform cylindrical AuNR array (dimensions < 400 nm) via electroplating is challenging\cite{Vasilantonakis2015}. Therefore, optimizing area-dependent electroplating parameters is highly necessary to fabricate a vertically aligned cylindrical AuNR array.

\section{Results}

\subsection{SEM analysis.} 
\begin{figure*}
    \centering
    \includegraphics[width=0.72\textwidth]{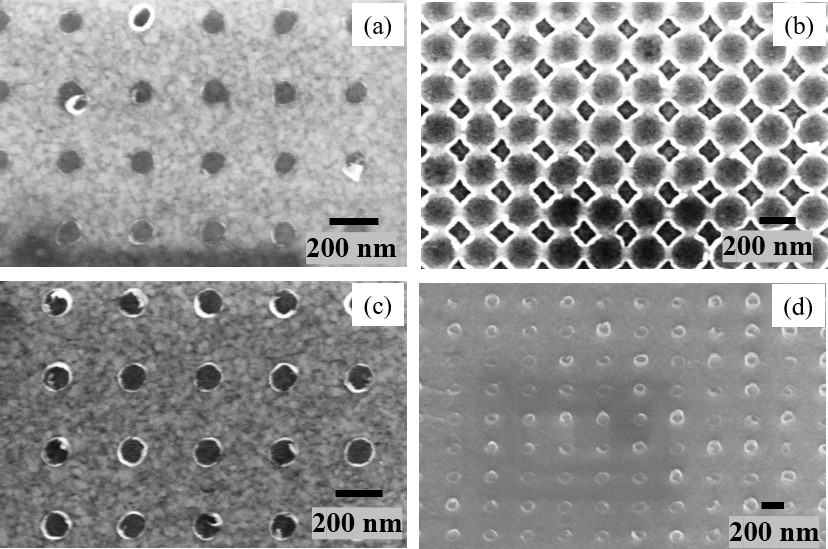}
    \caption{SEM images (surface-view) of vertically aligned AuNR array fabricated via electron beam lithography (EBL, 350 $\mu$C/cm\textsuperscript{2} dosage) and DC sputtering for various radius (r) and periodicity (p) combinations. (a) r = 40 nm, p = 200 nm, (b) r = 75 nm, p = 100 nm, (c)  r = 50 nm, p = 250 nm, (d) Tilted-view of SEM image (c).}
    \label{FIG:1}
\end{figure*}

Visualization of the vertical AuNR array is vital to measure its physical parameters. Fig.~\ref{FIG:1} shows the surface view of SEM images of a vertically aligned AuNR array formed on the Au thin-film deposited quartz substrate. 

A Ti/Au deposited thin film on a quartz substrate undergoes bilayer PMMA (A2/A4) spin coating to form a uniform 100 nm height. Masking or patterning of desired NR array was carried out for various EBL dosages ranging from 350 $\mu$C/cm\textsuperscript{2} to 400 $\mu$C/cm\textsuperscript{2} at different radius (r) and periodicity (p) combinations. The 350 $\mu$C/cm\textsuperscript{2} EBL dosage fabricates a uniform NR array; hence, it is considered for further fabrication. Fig.~\ref{FIG:1} (a) shows the surface view of the AuNR array for the 40/200 nm radius/periodicity combination. However, for the 75/100 nm masking combination, the AuNR merged at each edge of the array, as shown in Fig.~\ref{FIG:1} (b). A combination of 50 nm radius and 250 nm periodicity (center-to-center distance) helps to fabricate a uniform AuNR array compared to others, as shown in Fig.~\ref{FIG:1} (c). However, the tilt-view of the SEM image shows no sign of AuNR, i.e., broken or not deposited AuNR array, as confirmed by Fig.~\ref{FIG:1} (d). One reason for missing NR may be the limitation of small feature deposition in DC sputtering. Generally, for a multi-target sputtering chamber, the target, i.e., Au and the substrate, face each other with some slanting angle (10°-15°). After plasma formation between the target and substrate inside the sputtering chamber, the deposition of Au may partially block the edges of the patterned area (each 50 nm radius edge), which disallows for further deposition of Au in the hole. The deposited Au at the edges may lift off during PMMA removal, which can be verified by Fig.~\ref{FIG:1} (d), which shows a partially deposited and broken NR array.

\begin{figure*}
    \centering
    \includegraphics[width=0.85\textwidth]{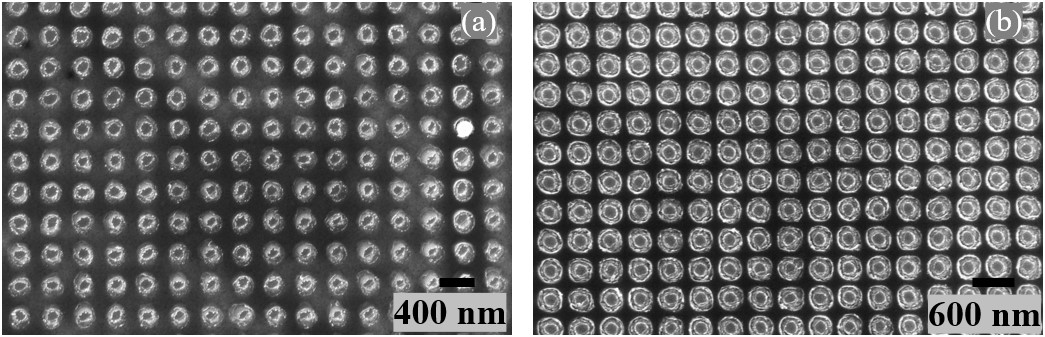}
    \caption{SEM images (surface-view) of vertically aligned AuNR array fibricated via EBL (at 350 $\mu$C/cm\textsuperscript{2} dosage) for (a) trilayer PMMA (A2/A4/A4) and (b) bilayer PMMA (A2/A4), and e-beam evaporation (r = 75 nm and p = 350 nm combination).}
    \label{FIG:2}
\end{figure*}

An e-beam evaporation technique was adopted to deposit a vertically aligned AuNR array to avoid slanting Au deposition. In e-beam evaporation, the desired material particles deposit perpendicularly to the substrate. Fig.~\ref{FIG:2} (a) and Fig.~\ref{FIG:2} (b) show the SEM images (surface view) of the AuNR array deposited on tri-layer PMMA (A2/A4/A4) and bilayer (A2/A4) PMMA, respectively, at 350 $\mu$C/cm\textsuperscript{2} EBL dosage for 75/350 nm radius/periodicity masking combination. A trilayer spin-coated PMMA forms a 170 nm thick layer, while the bilayer spin-coated PMMA forms a 100 nm thick layer, which leads to two distinct AuNR array formations. For the same development recipe, a bilayer spin-coated PMMA develops more than the trilayer PMMA, as shown in Fig.~\ref{FIG:2} (a) (the base or bottom area of the AuNR array). The bilayer spin-coated PMMA allows uniform deposition compared to the trilayer spin-coating, as shown in Fig.~\ref{FIG:2} (b). 

\begin{figure*}
    \centering
    \includegraphics[width=0.85\textwidth]{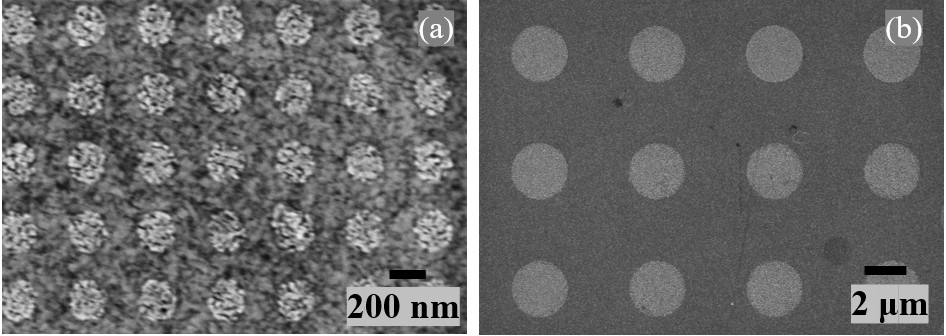}
    \caption{SEM images (surface-view) of vertically aligned AuNR array fabricated via EBL and electroplating for (a) Mask-I (r = 75 nm, p = 400 nm), (b) Mask-II (r = 1 $\mu$m, p = 5 $\mu$m).}
    \label{FIG:3}
\end{figure*}

A lower deposition rate (0.3-5 nm/min) in the e-beam evaporation technique limit to a deposit of higher Au thickness (>100 nm), as the vapour deposition process in vacuum heat up the instrument. Another well-established chemical deposition technique, electroplating, deposit a higher thickness of Au at a lower cost and less time in an ambient atmosphere. Fig.~\ref{FIG:3}  shows the SEM images (surface view) of a vertically aligned AuNR array for two different masks. The electroplating parameters for the said depositions are 0.002 A constant DC and 15 sec deposition time. The discussion section mentions a detailed optimization of the AuNR array via electroplating.

\subsection{AFM analysis.}

\begin{figure*}
    \centering
    \includegraphics[width=0.5\textwidth]{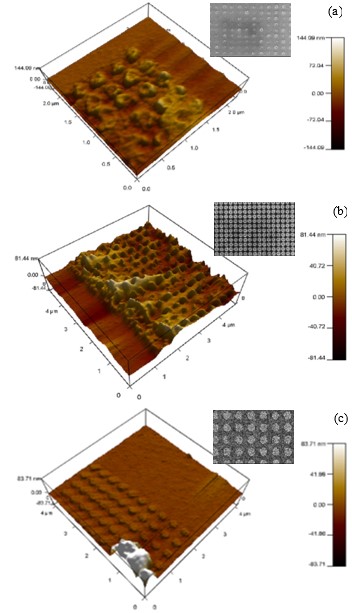}
    \caption{AFM image (3D-view) of vertically aligned AuNR array fabricated via EBL and (a) DC sputtering (b) e-beam evaporation, (c) electroplating. Inset images show corresponding SEM images.}
    \label{FIG:4}
\end{figure*}

The 3D AFM image analysis of the AuNR array highlights the geometry of the AuNR array fabricated using DC sputtering (Fig.~\ref{FIG:4} (a)), e-beam evaporation (Fig.~\ref{FIG:4} (b)), and electroplating (Fig.~\ref{FIG:4} (c)). Metallization via DC sputtering forms broken or partially deposited AuNR array, as discussed in the previous section, which is also confirmed in Fig.~\ref{FIG:4} (a). Conical-tip-shaped AuNR array was formed instead of a cylindrical one in e-beam evaporation, as shown in the 3D AFM image, i.e., Fig.~\ref{FIG:4} (b). However, a well-ordered, vertically aligned cylindrical AuNR array was formed in electroplating, as shown in Fig.~\ref{FIG:4} (c). 

\section{Discussion}

Both sputtering and e-beam evaporation techniques are costly and time-consuming for higher thickness (> 100 nm) deposition due to the heating up of the instrument to maintain the vacuum environment. However, the chemical deposition technique, i.e., electroplating, takes place in an ambient atmosphere, which is inexpensive and less time-consuming for Au deposition. The Au deposition depends on the seed-layer area, current density, and deposition time. As an area-dependent deposition method, the electroplating deposit Au only on the seed layer, unlike other vapor deposition techniques that deposit Au on the whole substrate, which makes the electroplating advantageous over others while removing the PMMA layer using acetone.

\begin{figure*}
    \centering
    \includegraphics[width=0.75\textwidth]{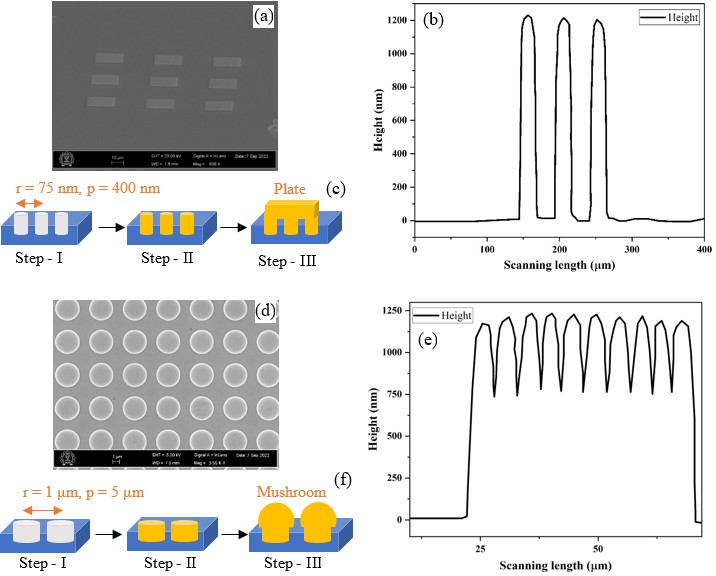}
    \caption{Mask-I (a) SEM image, (b) stylus profilometry measurement, (c) steps involved during Au deposition; Mask-II (d) SEM image, (e) stylus profilometry measurement, (f) steps involved during Au deposition via electroplating. Step-I: AuNR array pattern formed via EBL, Step-II: In-between Au deposition process, Step-III: final structure formed in electroplating.}
    \label{FIG:5}
\end{figure*}

To begin with, an Au thin film layer was deposited on the Ti/Au deposited quartz substrate (area, A = 0.6 cm\textsuperscript{2}) at a constant DC supply of 0.002 A (I) for 20 minutes (t). The stylus profilometry measurement gives an average thickness of 1400 nm (Supplementary Fig. S1), which matches the theoretical calculation, as described in equation 1. i.e., 1.40 $\mu$m. The uniform thin film deposition throughout the sample surface area makes it interesting to investigate for further study. 

\begin{equation}
    \text{Thickness} = (3.5275*10^4)\frac{I}{A} t
\end{equation}
In the next step, two mask structures, i.e., (i) Mask-I: r = 75 nm and p = 400 nm, and (ii) Mask-II, r = 1 $\mu$m and p = 5 $\mu$m, were fabricated using EBL (on bilayer PMMA, A2/A4, 100 nm thickness) followed by electroplating at a constant DC supply of 0.002 A for 60 sec. Fig.~\ref{FIG:5} (a) - Fig.~\ref{FIG:5} (c) show the SEM images, thickness measurement, and process involved for Mask-I. The SEM image shows plate formation instead of an NR array whose thickness is 1100 nm, as confirmed from surface profiler measurement (Fig.~\ref{FIG:5} (b)). Over time, the Au deposited in the pores quickly bridged on the top to form a plate-like structure due to smaller periodicity (i.e., 400 nm), as described in Fig.~\ref{FIG:5} (c). An intermediate stage of bridge formation among the filled pores is shown in supplimentary figure (S2). Similarly, for Mask-II (Fig.~\ref{FIG:5} (d) - Fig.~\ref{FIG:5} 5 (f)), mushroom structures were formed instead of the NR array, whose height was 1050 nm. In Mask-II, during Au deposition, pores were filled, followed by a uniform cap formed on the top (like a mushroom) of each NR due to larger periodicity (i.e., 5 $\mu$m), as described in Fig.~\ref{FIG:5} (f).

Immersion deposition also plays a vital role in depositing Au through gold sulfite solution because of its slightly acidic nature. Unknowingly, the sample was kept in the gold sulfite solution for five minutes after electroplating, which deposits Au through immersion deposition. Therefore, theoretical height deviates from the measured Au thickness. Uniform Au deposition for both miniaturized masks makes it further interesting to proceed with further investigation. As the EBL was carried out on 100 nm thick bilayer PMMA, which may allow plate-like Au deposition, and to avoid such undesired structure formation (e.g., plate), a higher thickness of PMMA was essential. Therefore, a monolayer PMMA (950K, 2\%) was spin-coated at a lower speed, 1000 rpm, forming a 1200 nm thick layer on the desired substrate. 

In the third phase of vertical AuNR array optimization, the same masks (Mask-I and Mask-II) were considered at 0.002 A DC for 15 sec of electroplating time, as shown in Fig.~\ref{FIG:3}. The sample was collected within 45 seconds after electroplating to avoid immersion deposition. Preliminary results show a uniform ordered AuNR array was formed for both the masks without any sign of plate-like (Mask-I) and mushroom-like (Mask-II) structure; however, the cylindrical top AuNR array possesses a porous NR surface due to lower thickness. AFM characterization, i.e., Fig.~\ref{FIG:4} (c), measures the height of the AuNR array deposited via electroplating to be 8.96 nm, and the roughness of a single AuNR top surface is 4.0 nm. Therefore, the lesser height of AuNR deposition might form a porous-like top surface, which could be improved by tuning the electroplating deposition parameters.

\begin{figure*}
    \centering
    \includegraphics[width=0.98\linewidth]{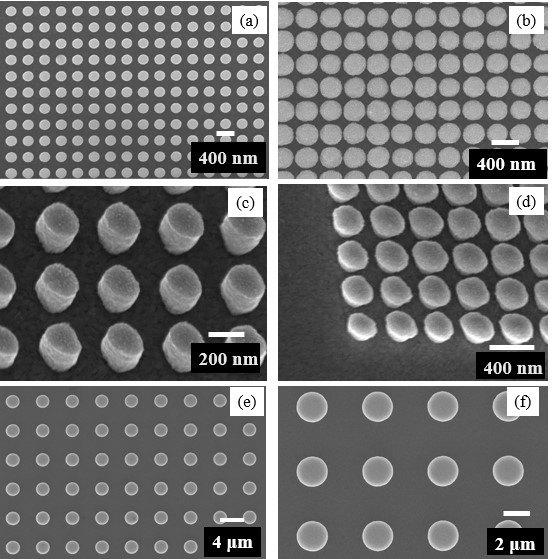}
    \caption{SEM images of vertically aligned AuNR array (Mask-I) fabricated via electroplating for EBL dosage at $\mu$C/cm\textsuperscript{2} (a) surface view, (c) titled-view (27°), and at $\mu$C/cm\textsuperscript{2} (b) surface view, and (d) titled-view (27°). SEM images of vertically aligned AuNR array (Mask-II) fabricated via electroplating for EBL dosage at (e) 450 $\mu$C/cm\textsuperscript{2} and (f) 500 $\mu$C/cm\textsuperscript{2}.}
    \label{FIG:6}
\end{figure*}

In the next phase, the electroplating current was reduced to 0.001 A to study the geometry of the vertical AuNR array. The electroplating deposition was carried out for 60 seconds and immediately taken out (within 10 seconds) to avoid immersion deposition in a gold sulfite solution. Fig.~\ref{FIG:3} shows smooth, uniform, and ordered AuNR array was fabricated via electroplating. Fig.~\ref{FIG:6} (a) and Fig.~\ref{FIG:6} (c) show the SEM images, surface view, and tilted view (27°) of vertically aligned cylindrical AuNR array fabricated for Mask-I at 450 $\mu$C/cm\textsuperscript{2} EBL dosage. Similarly, Fig.~\ref{FIG:6} (b) and Fig.~\ref{FIG:6} (d) show the SEM images, surface view, and tilted view (27°) of vertically aligned cylindrical AuNR array fabricated for Mask-I at 500 $\mu$C/cm\textsuperscript{2} EBL dosage. At lower EBL dosage, perfectly cylindrical AuNR array was formed whose dimensions were uniform throughout the whole array. However, the tilted view of the AuNR array formed at higher EBL dosage shows a cap-like or mushroom-like top structure on the top AuNR array.  Fig.~\ref{FIG:6} (e) and Fig.~\ref{FIG:6} (f) show the SEM images of a vertically aligned cylindrical AuNR array fabricated for Mask-II at 450 $\mu$C/cm\textsuperscript{2} and 500 $\mu$C/cm\textsuperscript{2}, respectively. The inset of Fig.~\ref{FIG:6} (e) shows the corresponding tilted view of the AuNR array. The AFM characterization (Supplementary Fig. S3) measures the height and roughness of each AuNR top surface of the vertical AuNR array to be 175 nm and 2.5 nm, respectively.

\newpage
In a vertically aligned AuNR array, the filling ratio $(\pi(r⁄p)^2)$ plays a vital role in detecting lower molecular weight analytes. A filling ratio between 0.35 to 0.6 is favorable for minute changes in the analyte, i.e., it detects a change in refractive index for the narrow peak shift wavelength in the visible wavelength. Table ~\ref{Table:1} accumulates actual design parameters for both the masks, EBL dosages, and obtained AuNR array geometry. A lower filling ratio for Mask-II may not be suitable for operating the optical biosensor in the visible region. However, a higher filling ratio for a miniaturized mask (Mask-I) for different EBL dosages is suitable for optical biosensors in the visible range.

\begin{table*}
\renewcommand{\arraystretch}{1.0}
\caption { Actual masks designed and dosage given in EBL, and diameter and spacing measured from SEM images in Fig. 6}
\begin{tabular}{|c|c|c|c|c|}
\hline
\textbf{EBL Mask Design} &
  \textbf{EBL Dosage ($\mu$C/cm\textsuperscript{2})} &
  \textbf{Measured diameter (nm)} &
  \textbf{Measured spacing (nm)} &
  \textbf{Filling ratio} \\ \hline

\begin{tabular}[c]{@{}c@{}} \textbf{Mask-I (r=75 nm, p=400 nm)}
\end{tabular}
& 450 &
  250 &
  400 &
  0.31 \\ \cline{2-5} 
 & 500 &
  340 &
  400 &
  0.57 \\ \hline
\begin{tabular}[c]{@{}c@{}} \textbf{Mask-II (r=1 $\mu$m, p=5 $\mu$m)}
\end{tabular} &
  450 &
  2010 &
  5000 &
  0.13 \\ \cline{2-5} 
 &
  500 &
  2192 &
  5000 &
  0.15 \\ \hline
\end{tabular}
\label{Table:1}
\end{table*}

\newpage
SEM and AFM characterizations confirm optimized vertically aligned cylindrical AuNR array fabricated via EBL and electroplating. Miniaturized AuNR array can be optimized for all combinations of radius and periodicity to get an optimal sensitive biosensor. Operating at a lower constant DC supply for a lesser possible immersion time in electroplating could lead to a smooth and ordered AuNR array rather than operating at a higher DC.

\section{Conclusion}

The higher cost and longer time required for AuNR array deposition (either via DC sputtering or e-beam evaporation) give birth to go for an easy and reliable electroplating deposition. Instrument heating during the deposition in a vacuum makes the physical vapour deposition method limited to deposit of less than 100 nm thickness. However, the electroplating deposition in the ambient atmosphere can deposit more than 100 nm of Au in a few seconds at a lower cost. The area-dependent electroplating deposition can tune the thickness depending on the deposition time and DC supply. Slightly acidic (6.0 < pH < 7.0) gold sulfite solution impact adversely the Au thickness due to immersion deposition. The immersion deposition may form plate-like (for lesser radius and spacing combination) and mushroom-like (for larger radius and spacing combination) structures and, therefore, should be as minimum as possible. Depositing miniaturized Au structure via electroplating at a lower DC supply is beneficial compared to deposition at a higher DC supply. A higher filling ratio between 0.35 and 0.6 using Mask-I is suitable for fabricating the optical biosensor in the visible region. 

\section{Methods}

\begin{figure*}
    \centering
    \includegraphics[width=0.89\textwidth]{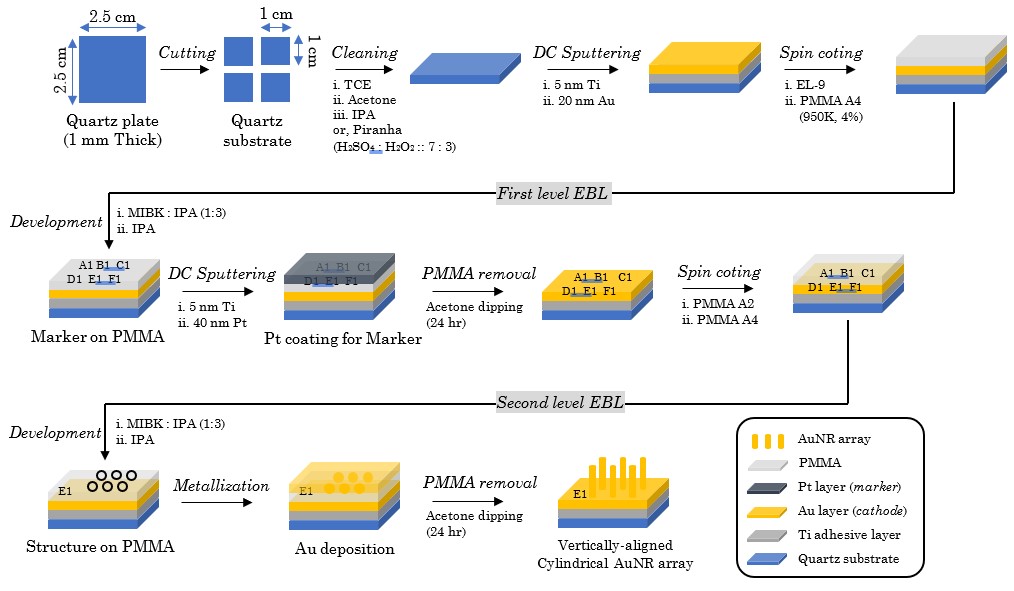}
    \caption{Schematic of an experimental procedure to fabricate a vertical AuNR array via electron beam lithography (EBL) and metallization.}
    \label{FIG:7}
\end{figure*}

Fig.~\ref{FIG:7} shows the schematic for experimental procedures involved in fabricating vertical Au NR array via EBL and metallization. One-millimeter-thick quartz substrate (2.5×2.5 cm\textsuperscript{2}) was cut into 1×1 cm\textsuperscript{2} and cleaned using an ultrasonic sonicator in trichloroethylene, acetone, and isopropyl alcohol (IPA) solution (each for five minutes). Piranha cleaning (H\textsubscript{2}SO\textsubscript{4}: H\textsubscript{2}O\textsubscript{2}:: 7 : 3) was carried out for 30 minutes in an ambient atmosphere to remove sodium and potassium remnants. A five nm thin adhesive Ti-layer was deposited on the cleaned quartz substrate via DC sputtering (2.1 nm/min). A 20 nm Au thin film layer was deposited via DC sputtering (17 nm/min) on the adhesive Ti-layer to make a transparent and conducting quartz substrate. 

First level EBL (make/model: Raith GmbH/150-Two) was used to make a marker pattern on the substrate. For easy lift-off of the polymethyl methacrylate (PMMA A4: 950K, 4\%) resist layer, a co-polymer resist ethyl lactate (EL-9) layer (3200 rpm, 50 sec) was spin-coated on the Au layer (post-bake: 180$^{\circ}$C, 5 min) followed by PMMA layer (2000 rpm, 50 sec). After baking at 180$^{\circ}$C for two minutes, the marker pattern was written via EBL performed at 20kV, a dosage of 350$\mu$C/cm$^2$  beam aperture of 60$\mu$m with a current density of 1.1005 nA. Then the sample was first kept in methyl isobutyl ketone (MIBK) and IPA solution (3:1) for 30 seconds to develop the pattern and then put in IPA solution (40 seconds) to cease further development. Metallization was carried out in DC sputtering to deposit Ti/Pt (20 nm/40 nm) on the PMMA layer. The sample was kept in acetone for 15 hours to wipe out the PMMA layer and get the desired marked pattern. 

Second-level EBL was carried out to make the vertical AuNR array pattern on the sample. A bilayer PMMA (A2 and A4) was spin-coated on the marked sample after baking for one minute at 180℃. The first PMMA (A2: 950K, 2\%) layer was spin-coated (2500 rpm, 50 sec) on the sample and post-baked for two minutes at 180℃, then the second PMMA (A4: 950K, 4\%) layer was spin-coated (3000 rpm, 50 sec) on PMMA A2 layer and post-baked for two minutes at 180℃. Then the vertical AuNR array pattern was written on the PMMA layer at 20kV, a dosage of 350 $\mu$ C $cm^{-2}$, beam aperture of 30$\mu$m with a current density of 0.25619 nA. Then a similar procedure was followed to develop the pattern, as mentioned for marker formation. 

\subsection{Metallization via DC Sputtering.}

The sputtering instrument (AJA International, Inc/ Orion Sputter PHASE) was used to deposit Ti/Au layers on the patterned PMMA layer. DC sputtering was preferred over RF sputtering to deposit the Au layer due to the higher deposition rate, which disallowed the temperature to cross its limiting value. The substrate holder rotates at 10 rpm/min to deposit a uniform layer. A 20 nm thin Ti-layer was deposited as an adhesive layer on PMMA at a DC power of 150 watt and 3.0 mTorr pressure (deposition rate of 2.1 nm/min). Again, DC sputtering was operated to deposit a 150 nm thin Au layer on the Ti-layer at 100 watt DC power and 3.0 mTorr pressure (deposition rate of 17 nm/min). Then the sample was kept in an acetone solution overnight to lift off the PMMA layers.

\subsection{Metallization via E-Beam Evaporation.}

The e-beam evaporation instrument (AJA International/ ATC-ORION-8E) was used to deposit Ti/Au layers on the patterned PMMA layer of the sample. The substrate holder rotates at 10 rpm/min to deposit a uniform layer. A 20 nm thin Ti-layer was deposited as an adhesive layer on PMMA at a 0.3 nm/min deposition rate. Again, e-beam evaporation was operated to deposit a 100 nm thin Au layer on the Ti-layer at a 5 nm/min deposition rate. Then the sample was kept in an acetone solution overnight to lift off the PMMA layers.

\subsection{Metallization via Electroplating.}

\begin{figure}
    \centering
    \includegraphics[width=0.4\textwidth]{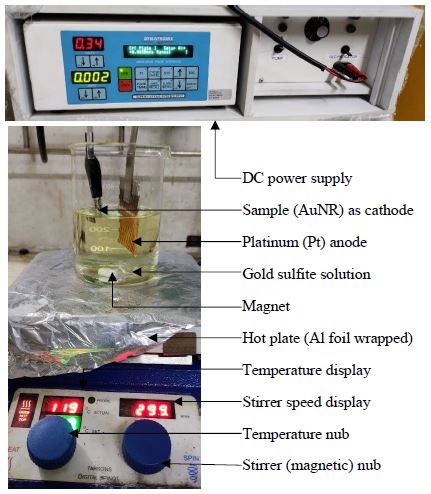}
    \caption{Experimental set-up for Au deposition via electroplating.}
    \label{FIG:8}
\end{figure}

Au electroplating was carried out on a hot plate at 120℃ with a magnetic stirrer speed of 300 rpm (as shown in Fig.~\ref{FIG:8}) to avoid chunk formation. The temperature of the measured gold sulfite solution (TSG-250, one troy ounce of gold per gallon) (kept in the beaker) was 50℃. In the chemical deposition method, platinum (Pt) metal was used as an anode, and Au deposited quartz substrate (sample) was dipped inside the gold sulfite solution as a cathode. The DC power supply was operated in constant current mode at 0.002 A DC for 15 sec. As the gold sulfite solution is slightly acidic/neutral (pH: 6.0-7.0), the sample was immediately taken out of the solution and dipped in DI water to avoid immersion deposition. 

\newpage
\section{Characterization}
SEM (make: Raith GmbH/ 150-Two), AFM (make: Asylum/Oxford Instruments, MFP3D Origin), and surface profiler (make: Bruker/ Dektak XT) characterizations were performed to measure geometry, and roughness of the fabricated cylindrical vertical AuNR array. 

\bibliography{sample}

\noindent

\section*{Acknowledgements}

A.K. acknowledges funding support from the Department of Science
and Technology via the grants: SB/S2/RJN-110/2017,
ECR/2018/001485 and DST/NM/NS-2018/49. We acknowledge the Centre of Excellence in Nanoelectronics (CEN) at IIT Bombay for providing fabrication and characterization facilities.  M.K. acknowledges Institute Postdoctoral fellowship for financial support. A.A. acknowledges UGC for his junior research fellowship to support PhD.

\section*{Author contributions statement}

A. K. and M.K.S. conceived the idea. M.K.S., A.A., and A.K. designed the research. A.A. performed EBL experiment. M.K.S. performed metallization experiments and analyzed the data on the manuscript. M.K.S., A.A., and A.K. discussed the results. M.K.S. wrote the manuscript. All authors reviewed the manuscript.

\end{document}